# Origin of ferroelectric-like hysteresis loop of $CaCu_3Ti_4O_{12}$ ceramic studied by impedance and micro-Raman spectroscopy

Sungmin Park, Hyosang Kwon, Doyoung Park, Hyeonsik Cheong, Gwangseo Park*

*Department of Physics, University of Sogang, Seoul, Korea*

## ABSTRACT

Ferroelectric-like hysteresis loops of $CaCu_3Ti_4O_{12}$ (CCTO) ceramic have been observed. We found that this unusual feature does not arise from the displacement of the Ti ions in the $TiO_6$ octahedron, but apparently comes from the charges at the grain boundaries which consist of a CuO layer. The relaxation time of 2.9 milliseconds by the charges from the grain boundary, nearly corresponding to the inverse $P - V$ sampling frequency of 1kHz, has been found in the impedance spectrum. According to the micro-Raman mapping, the CuO layer is found in the grain boundary and is perfectly distinguished from the CCTO grain.

# INTRODUCTION

The cubic perovskite-related structure (space group $Im3$) $CaCu_3Ti_4O_{12}$ (CCTO) has a colossal dielectric constant (CDC) of the order $10^5$, which remains nearly constant from 100 K up to 400 K but drops rapidly down to less than 100 below 100 K. [1-3] Even though several studies have been made in an attempt to understand the origin of intrinsic or extrinsic colossal dielectric constant, they are not still widely acceptable. Besides debates regarding the colossal dielectric constant, the ferroelectricity of CCTO is also an important issue, yet it has not been thoroughly investigated. Liu *et al*. studied its intrinsic nature and demonstrated that the origin of ferroelectricity is strongly correlated with one dimensional <001> columns of $TiO_6$ octahedra, suggesting that the structurally frustrated-relaxor ferroelectric behavior of CCTO plays a significant role in deciding the colossal dielectric constant.[4] Presenting a different point of view, Prakash *et al*. suggested that local electrical polarization induced by the simultaneous presence of $Ti^{3+}$ and $Ti^{4+}$ forming an asymmetrical bonding of $Ti^{3+}$-O-$Ti^{4+}$ is the reason behind the hysteresis loop of CCTO ceramic. [5] Although this is a persuasive argument, it is unclear how the charges generated from the presence of $Ti^{3+}$ and $Ti^{4+}$ could move through the grains and grain boundaries and reach the electrodes while the ac voltage for sampling the $P - V$ curve is being applied.

In this report, using micro-Raman spectroscopy, we demonstrate what type of materials form the grain and grain boundary, which has been being an important issue related to the electrical properties of CCTO ceramic sintered at over 1100 °C. In addition, using impedance spectroscopy, it will be demonstrated how the charges from the grain boundary can produce the hysteresis loop.

# EXPERIMENTS

Ceramic samples were prepared by the conventional solid-state reaction technique using starting materials of $CaCO_3$ (Aldrich, 99.995%), CuO (Aldrich, 99.99%) and $TiO_2$ (Aldrich, 99.99%). The raw materials were weighed according to their stoichiometric ratio and mixed thoroughly in an agate mortar. The mixture was pressed into pellets of 10-mm diameter with 0.5-mm thickness, and they were then sintered in various temperature ranges in air for 12 hours, before being ground and polished carefully as small disks. Small patches of conductive silver paste were then coated onto the top surface of the disks to form electrodes with a separation of 0.4 ~ 1 mm, followed by being heated at 50 °C for 3 hours to dry the residual solvent in the silver paste and obtain good electrical contact.

The crystal structures were identified by x-ray diffraction using Bruker D8 DISCOVER with Cu-$K\alpha$ radiation of 1.5406 Å, and by Raman spectroscopy at room temperature. For the micro-Raman spectra, a 514.5-nm line $Ar^+$ ion laser with a power of 1 mW was used. The laser beam was focused onto the sample by a 40× microscope objective lens (0.6 N.A.), and the scattered light was collected and collimated by the same objective lens. The scattered light was filtered with a holographic edge filter, dispersed by a Spex 0.55-m spectrometer, and detected with a liquid-nitrogen-cooled back-illuminated charge-coupled-device detector array. The spatial resolution was about 500 nm, and the spectral resolution was about 1 $cm^{-1}$.

Precision LC (RADIANT) with a virtual ground mode and HP4194 impedance analyzer were used to investigate the electrical properties such as $P - V$ (polarization *vs*. voltage) and $C - V$ (capacitance *vs*. voltage) curves.

# RESULTS AND DISCUSSION

CCTO ceramic has prominent x-ray diffraction peaks at (220), (400) and (422) which correspond exactly to those of CCTO in the JCPDS file without secondary phases. From the macro-Raman spectra obtained at room temperature, the peaks observed at 446, 511 and 574 $cm^{-1}$ are quite well matched with those of the theoretical values of $A_g(1)$, $A_g(2)$ and $F_g(3)$ modes in CCTO, respectively. [6] It is known that $A_g(1)$ and $A_g(2)$ modes are related to $TiO_6$ rotation-like atomic motion, and $F_g(3)$ originates from O-Ti-O anti-stretching atomic motion. [6]

The grain size of CCTO ceramic is widely distributed in the range from 5 to 60 μm, whereas the thickness of the grain boundary is between 1.0 and 5.8 μm, which is larger than those in previous reports. The grain size and the thickness of the grain boundary are estimated by taking over 50 regions from a FE-SEM image.

Figure 1 show the $P-V$ hysteresis loop measured at room temperature for the CCTO pellet doubly sintered at 1100 °C ('doubly' means that the ceramic has been sintered two times in the same way). Two models have generally been proposed to explain a hysteresis loop in CCTO. Prakash *et al*. claims that the simultaneous presence of $Ti^{3+}$ and $Ti^{4+}$ forming an asymmetrical bonding of $Ti^{3+}$-O-$Ti^{4+}$ can induce locally driven electrical polarization. [5] On the other hand, considering an intrinsic origin of the ferroelectricity, it has been demonstrated that in the absence of an electric field, Ti ions in $TiO_6$ octahedra can be displaced along <001> columns like a common ferroelectric material such as $BaTiO_3$. [4] However, according to Bozin *et al*., the octahedral tilt angle must be less than 0.3° at all temperatures, clearly implying very little displacement of the Ti ion due to the immovable Cu-O bond length. [7] Consequently, the local disorder (particularly the tilt of the O-Ti-O bond angle) or the off-center Ti displacements must be not significant enough even under a higher static electric field. As a result, it should be concluded that the charges contributing to hysteresis-like $P - V$ loop cannot originate from an intrinsic origin, but rather from extrinsic factors such as defects or vacancies. In order to investigate which charges in the CCTO ceramic would be exactly related to the sampling frequency of ~1 kHz during the $P - V$ curve measurement, we carried out impedance spectroscopy. In principle, by introducing dielectric relaxation, impedance spectroscopy can give dielectric and conduction information regarding mobile and ionic charges, depending on the sampling frequency. [8] After being precisely fitted for the impedance spectrum at 300 K, it is found that the equivalent circuit consists of four parallel $RC$ elements in series, as shown in the inset of Fig. 2(b). $R_g$, $C_g$, $R_{gb}$ and $C_{gb}$ denote the resistance and the capacitance of the grain, the resistance and the capacitance of the grain boundary, respectively (1 and 2 denote the first and second grain or grain boundary, respectively). For two parallel $RC$ elements in series for example, one can find the real (Z′) and imaginary (Z″) part of impedance ($Z^*$) as a function of the sampling frequency as shown in Fig. 2(a). The impedance $Z^*$ for the equivalent circuit in Fig. 2(a) is generalized as

$$Z^* = \frac{1}{R_1^{-1} + i\omega C_1} + \frac{1}{R_2^{-1} + i\omega C_2}. \tag{1}$$

where

$$Z' = \frac{R_1}{R_1 + (\omega R_1 C_1)^2} + \frac{R_2}{R_2 + (\omega R_2 C_2)^2} \tag{2}$$

And

$$Z'' = R_1\left(\frac{\omega R_1 C_1}{1+(\omega R_1 C_1)^2}\right) + R_1\left(\frac{\omega R_2 C_2}{1+(\omega R_2 C_2)^2}\right). \tag{3}$$

Two highest peaks of $Z''$ occur at frequencies of $\omega_1=(R_1C_1)^{-1}$ and $\omega_2=(R_2C_2)^{-1}$, where each *RC* has a relaxation time $\tau$ giving us a characteristic time of certain charges, and $R_1$ is larger than $R_2$. In our case, even though it seems to be difficult to identify some of the peaks in $Z''$ at 300 K, the fitting result could provide us with the presence of three kinds of relaxation time (originally, four kinds of relaxation time should be obtained due to four parallel *RC* elements in series. However, one of $Z''$ peaks is already moved forward to a higher frequency region more than ~ 1 MHz, thus we could not obtain it). From the fitted values of the impedance spectrum at 300 K, it is found that the relaxation time of ~ 2.9 milliseconds observed at the frequency of ~ 345 Hz is clearly attributed to the grain boundary in the CCTO ceramic due to large resistance and capacitance of the grain boundary compared to those of the grain. This relaxation time is not a perfect match with the inverse sampling frequency for the *P - V* measurement, but is reasonable enough to be related to the main relaxation around 1 kHz if the non-ideal dielectric relaxation is taken into account (we fitted the impedance spectra with the Cole-Cole equation which regards the relaxation process as non-ideal). This interpretation suggests that the charges giving rise to the relaxation time of 2.9 milliseconds should contribute to the *P - V* curve measured at 1 kHz.

In previous reports on the grain and its boundary, it was found that the insulating barrier layer at the grain boundary could be formed by a reoxidation process during cooling down to room temperature, due to the limited oxidation of the grain boundary. [9] According to energy dispersive x-ray (EDX) analyses on the insulating layer, this region has been mostly believed to be Cu rich or have CuO-related layers. [10, 11] Especially, Adams *et al*. [11] found that CCTO ceramic pellets sintered at 1000°C under ambient $N_2$ or $O_2$ decompose into a two-region phase mixture, such as $CaTiO_3$, $Cu_2O$ or $TiO_2$. However, the x-ray dispersive-based techniques cannot give us a straightforward interpretation of the crystal structure and its symmetry within the local region. So instead we have used micro-Raman spectroscopy to identify the sort of elements present in the grains and grain boundaries. The micro-Raman scanning was carried out in a 20μm×20μm region (the spatial resolution is ~ 500 nm). From two maps of the micro-Raman spectrum shown in Fig. 3, one can find that the grains and the grain boundaries consist of CCTO and CuO, respectively (the presence of the CuO is not found in the XRD patterns due to the overlaps with CCTO). The grains (light red) and grain boundary regions (black) have been perfectly separated and the image is exactly comparable to the optical one (inset, Fig. 3(a)). Figure 3(a) shows the Raman image for 511 cm$^{-1}$ peak intensity corresponding to the $A_g(2)$ mode of CCTO. In addition, the 511 cm$^{-1}$ peak intensities inside the grain show no significant variation, which indicates homogeneous growth of the CCTO grain. Figure 3(b) shows the Raman intensity image for 347 cm$^{-1}$ corresponding to the Raman active mode of CuO single crystal. [12] Although the grain boundary for CCTO ceramic sintered at more than 1100 °C has, according to conventional EDX analysis, been being believed to be composed of $CaTiO_3$, CuO, $Cu_2O$ or $TiO_2$, a direct structural understanding as we show here, has never been done before. This finding is quite consistent with the early suggestion of Romero and Leret *et al*. [13, 14]

CuO is generally considered to be a Cu deficient *p*-type semiconductor with a band gap of ~ 1.4 eV [15]. The semiconductivity of CuO originates from the presence of $Cu^{3+}$ ions, commonly identified by the XPS spectrum of Cu $2p_{3/2}$. [16] If the charges contributing to the $P - V$ curve arise from the CuO, then its activation energy could be close to that of single crystal or ceramic CuO. According to recent results, the activation energy for CuO grain at room temperature is found to be about 0.10 ~ 0.23 eV including that of the grain boundary. [17 ~ 20] However, for our sample, the activation energy of the CuO-based grain boundary, extracted from the conductance and relaxation time, is 0.50 ~ 0.60 eV, which is similar to that of the CCTO grain boundary in previous reports. At this point, we cannot definitely conclude why they do not show similar activation energy.

## CONCLUSION

In summary, by using micro-Raman spectroscopy, we found that the grain boundary region in the CCTO ceramic sintered at 1100 °C consists of a CuO layer, and by using impedance spectroscopy, its charges give rise to a relaxation time of ~ 2.9 milliseconds, contributing to the hysteresis-like $P$ - $V$ curve measured at 1 kHz. We also demonstrate that the combination of impedance and micro-Raman spectroscopy would be able to provide a new way for understanding intrinsic origin of electrical properties in a system made up of more than two kinds of oxide material.

## ACKNOWLEDGEMENTS

This research was supported by Basic Science Research Program through the National Research Foundation of Korea (NRF) funded by the Ministry of Education, Science and Technology (2011-0012372 and 2008-2004744).

# REFERENCES

1. M.A. Subramanian, D. Li, N. Duan, B. Reisner and A.W. Sleight, Solid State Chem. **151**, 323 (2000).

2. A.P. Ramirez, M.A. Subramanian, M. Gardel, G. Blumberg, D. Li, T. Vogt and S.M. Shapiro, Solid State Commun. **115**, 217 (2000).

3. Subramanian, D. Li, N. Duan, B. Reisner and A.W. Sleight, Solid State Chem. **151**, 323 (2000).

4. Yun Liu, Ray L. Withers, Xiao Yong Wei, Phys. Rev. B, **72**, 134104 (2005).

5. B. Shri Prakash and K. B. Varma, Appl. Phys. Lett. **90**, 082903 (2007).

6. N. Kolev, R. P. Bontchev, V. N. Popov, V. G. Hadjiev, A. P. Litvinchuk, M. N. Iliev, Phys. Rev. B, **66**, 132102 (2002).

7. E S Bozin, V Petkov, P M Woodward, T Vogt, S D Mahanti and S J L Billinge, J. Phys.: Condens. Matter **16**, 5091(2004).

8. Andrew K Jonscher, J. Phys. D: Appl. Phys. 32, 57(1999).

9. Derek C. Sinclair, Timothy B. Adams, Finlay D. Morrison, and Anthony R. West, Appl. Phys. Lett. **80**, 2153 (2002).

10. Tsang-Tse Fang and Li-Then Mei, J. Am. Ceram. Soc. **90**, 638 (2007).

11. Timothy B. Adams, Derek C. Sinclair, and Anthony R. West, J. Am. Ceram. Soc. **89**, 2833 (2006).

12. H. F. Goldstein, Dai-sik Kim, Peter Y. Yu, and L. C. Bourne, Phys. Rev. B, **41**, 7192 (1990).

13. J. J. Romero, P. Leret, F. Rubio-Marcos, A. Quesada, J. F. Fernández, J. Eur. Cer. Soc, **30**, 737 (2010).

14. P. Leret, J. F. Fernandez, J. de Frutos, D. Fernández-Hevia, J. Eur. Cer. Soc, **27**, 3901 (2007).

15. F. P Koffyberg and F. A. Benko, J. Appl. Phys **53**, 1173 (1982).

16. J. Ghijsen, L. H. Tjeng, J. van Elp, H. Eskes, J. Westerink, and G. A. Sawatzky, Phys. Rev. B **74**, 024106 (2006).

17. Yong kwon Jeong and Gyeong man Choi, J. Phys. Chem. Solids **57**, 81 (1996).

18. Ming Li, Antoni Feteira, and Derek C. Sinclair, J. Appl. Phys. **105**, 114109 (2009).

19. Sudipta Sarkar, Pradip Kumar Jana, and B. K. Chaudhuri, Appl. Phys. Letts **89**, 212905 (2006).

20. Sudipta Sarkar, Pradip Kumar Jana, and B. K. Chaudhuri, Appl. Phys. Letts **92**, 022905 (2008).

**FIGURES**

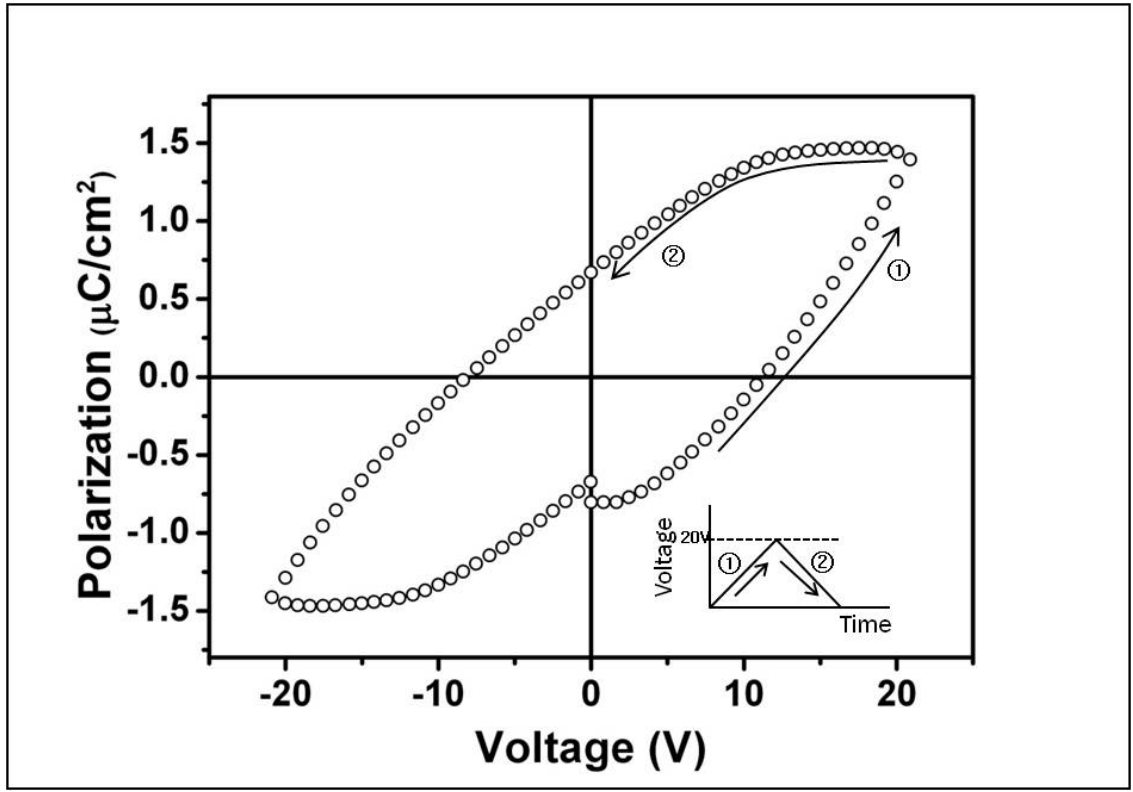

Figure 1. The $P - V$ hysteresis loop of the CCTO ceramics measured at 300K is shown. This CCTO ceramic is sintered twice at 1100 °C.

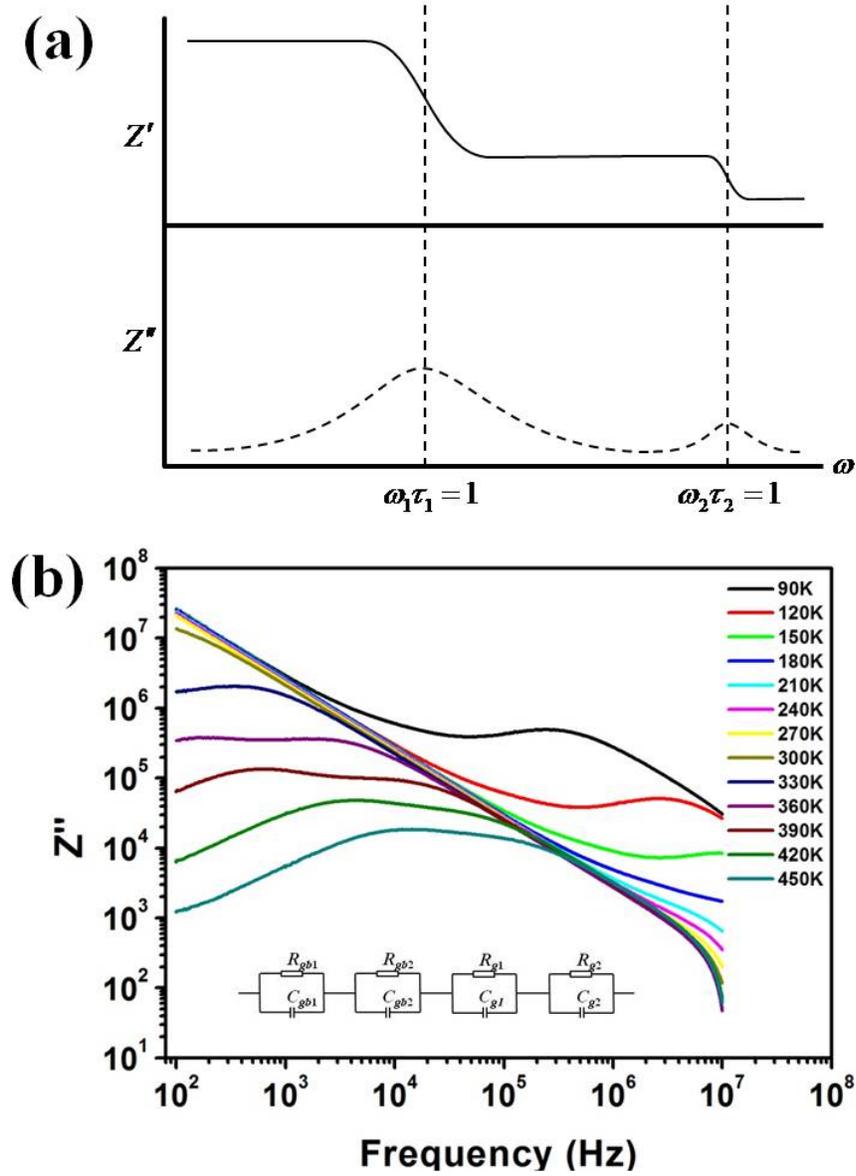

Figure 2. (a) The real (Z′) and imaginary part (Z″) of the impedance spectra in an ideal relaxation process for two parallel RC elements in series are shown. (b) The imaginary part of the impedance spectra (Z″) of the CCTO ceramic measured at various temperatures are shown here. The inset depicts the equivalent circuit consisting of four parallel RC elements in series.

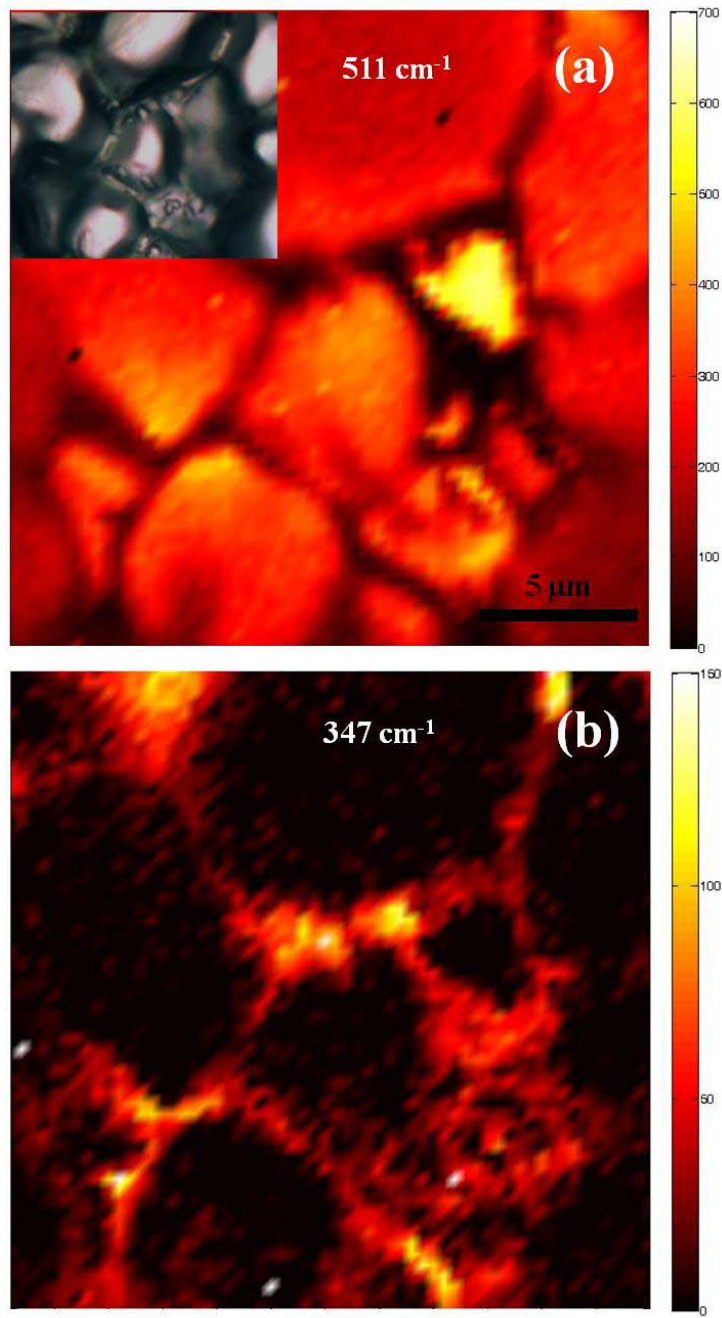

Figure 3. These figures show two different Raman intensity maps of (a) 511 cm$^{-1}$ of the CCTO and (b) 347 cm$^{-1}$ of the CuO. The inset of (a) shows the optical microscope image of the sample.